\documentclass{article}
\usepackage{amssymb}

\usepackage{epsfig}
\usepackage{graphicx}
\usepackage{amsmath}

\input{tcilatex} 

\begin{document}

\bigskip 
\begin{titlepage} 
\begin{center} {\LARGE \bf Studies of the Schroedinger-Newton Equations in $D$ Dimensions} 
\\ \vspace{2cm} R. Melko\footnotemark\footnotetext{email: 
rgmelko@gandalf.uwaterloo.ca}  and  R.B. Mann \footnotemark\footnotetext{email:  mann@avatar.uwaterloo.ca}  \\
 \vspace{1cm} Dept. of Physics, University of Waterloo Waterloo, ONT N2L 3G1, Canada\vspace{2cm} \today\\ \end{center}  
\begin{abstract}  
We investigate a $D$ dimensional generalization of the Schroedinger-Newton equations, which purport to 
describe quantum state reduction as resulting from gravitational effects.  For a single particle, 
the system is a combination of the Schroedinger and Poisson equations modified so that the probability 
density of the wavefunction is the source of the potential in the Schroedinger equation.
For spherically symmetric wavefunctions, a discrete set of energy eigenvalue solutions emerges for 
dimensions $D<6$, accumulating at $D=6$.  Invoking Heisenberg's uncertainty 
principle to assign timescales of collapse correspoding to each energy eigenvalue,
we find that these timescales may vary by many orders of magnitude depending on dimension.  
For example, the time taken for the wavefunction of a free neutron 
in a spherically symmetric state to collapse is many orders of magnitude
longer than the age of the universe,  whereas for one confined to a 
box of picometer-sized cross-sectional dimensions the collapse time is about two weeks.
\end{abstract} 
\end{titlepage} 

\section{Introduction}

In quantum mechanics, objects are described by wavefunctions. These take the
form of complex superpositions of various evolutionary alternatives, or
states. Although successful in describing many aspects of the quantum world,
this picture often leads to troubling interpretations when extrapolated to
the macroscopic level. One issue that has suffered long debate is the fact
that one never observes a superposition of states. Rather, one only observes
a system's \textit{basic} or \textit{stationary} states. We are therefore
forced to provide a mechanism by which quantum wavefunctions reduce to their
stationary states. This process is called \textit{wavefunction collapse} or 
\textit{state reduction}. Motivated by the basic conflicts which exist
between general relativity and quantum mechanics, a number of authors have
proposed the idea that wavefunction collapse is an objective phenomenon
which arises due to gravitational effects \cite{qgstate}. \ For example
Penrose \cite{penrose} has suggested a scheme in which a superposition of
two stationary quantum states should be fundamentally unstable if there
exists a significant mass displacement between them. In this case there
should be some characteristic timescale $T_{G}$ for decay into the basic
states. \ Although a detailed estimate of $T_{G}$ would require a full
theory of quantum gravity, under this hypothesis it is reasonable to expect
that for non-relativistic systems 
\begin{equation}
T_{G}\simeq \frac{\hslash }{\Delta E_{G}}  \label{e1}
\end{equation}
where $\Delta E_{G}$ is the gravitational self-energy of the difference
between the mass distributions of the two states.

The explicit nature of the basic states in this consideration is somewhat
unclear. We cannot simply regard the position of a lump of mass as a basic
state, because then we would be forced to regard any general state of a
particle as a superposition. As a possible solution to this problem, Penrose
proposes that these (non-relativistic) basic states are solutions of the
Schroedinger equation 
\begin{equation}
-\frac{\hslash ^{2}}{2m}\nabla ^{2}\Psi +U\Psi =E\Psi   \label{e2}
\end{equation}
where the additional term represents a coupling to a certain gravitational
potential $U$. This potential is determined (via the Poisson equation) by
the expectation value of the mass distribution in the state determined by
the wavefunction. For single particle systems, the matter density is
determined by the probability density from the wavefunction, and so 
\begin{equation}
\nabla ^{2}U=4\pi Gm^{2}\left| \Psi \right| ^{2}  \label{e3}
\end{equation}
where $G$ is Newton's gravitational constant, and $m$ is the mass of the
single particle. Equations (\ref{e2},\ref{e3}) are dubbed the
Schroedinger-Newton (SN)\ equations \cite{penrose2}. A\ preliminary
investigation of the properties of the solutions to the SN equations was
recently carried out by Moroz et. al. \cite{moroz}. Under the assumptions of
spherical symmetry in $3$ dimensions, and by demanding only that $U$ and $%
\Psi $ be everywhere smooth, they discovered a discrete family of bound
state solutions, labelled by an integer $n\geq 0$. Each solution is a
normalizable wavefunction, and the $n$th solution has $n$ zeros. The energy
eigenvalues associated with each of these solutions are negative, and
monotonically converge to zero for large $n$. \ These results can be
justified analytically \cite{todmor}. The energy eigenvalues are the
differences between a given bound state and a continuum `superposition'
state, and so provide via\ (\ref{e1}) an estimate of the timescale of
self-collapse of a single particle of mass $m$. The energy eigenvalues scale
like $m^{5}$, and so particles of small mass have extremely long
self-collapse times -- for a nucleon mass the estimate is $10^{53}$s \cite
{moroz}. A recent related study by Soni is commensurate with these results 
\cite{soni}.

Relaxing the assumption of spherical symmetry is in general a difficult task
due to the non-linearity of the SN equations. However there are two
situations in which this is fairly straightforward: cylindrical symmetry
with no angular momentum and planar symmetry. Rewriting the SN equations for
these cases effectively reduces them to $2$ and $1$ dimensional situations
respectively. These cases, along with the spherically symmetric case, can be
simultaneously recovered by rewriting the spherically symmetric SN equations
in $D$ dimensions. Motivated by the above, we consider in this paper an
analysis of the $D$-dimensional spherically symmetric SN equations, for $%
D\geq 1$. Although the higher-dimensional cases are of less direct physical
interest that the $D=2,3$ cases, such a study affords us some insight into
the dimensional behaviour of the SN\ system. This behaviour may be of more
than pure pedagogical interest since many candidate approaches to quantum
gravity are typically cast in higher dimensions (superstring theory being
the obvious example).

\section{The D-dimensional SN equations}

Any solution to the SN equations (\ref{e2},\ref{e3}) must be normalizable
(i.e. square-integrable). Integrating the probability density over all space
yields 
\begin{equation}
k^{2}=\int_{0}^{\infty }d^{D}x\,\left| \Psi \right| ^{2}  \label{norm1}
\end{equation}
where $k$ is a dimensionless number, and so the wavefunction must be
rescaled to ensure there is unit probability of finding the particle
somewhere in space. \ Writing \ $\Psi =k\psi $, the SN equations then become 
\begin{eqnarray}
-\frac{\hslash ^{2}}{2m}\mathsf{\nabla }^{2}\psi +U\psi &=&E\psi  \label{snA}
\\
\mathsf{\nabla }^{2}U &=&4\pi Gk^{2}m^{2}\left| \psi \right| ^{2}
\label{snB} \\
\int_{0}^{\infty }d^{D}\mathsf{x}\,\left| \psi \right| ^{2} &=&1  \label{snC}
\end{eqnarray}
and we see that the normalization factor enters the system due to its
non-linearity. Redefining variables in (\ref{snA},\ref{snB}) via \cite{moroz}
\begin{equation}
\psi =\alpha \mathcal{S}\text{ \ \ \ \ \ \ \ \ }E-U=\beta \mathcal{V}
\label{e4}
\end{equation}
where 
\begin{equation}
\alpha =\frac{\hslash }{\sqrt{8\pi Gk^{2}m^{3}}}=\frac{\hat{\alpha}}{k}\text{
\ \ \ \ \ \ \ \ }\beta =\frac{\hslash ^{2}}{2m}  \label{e5}
\end{equation}
yields 
\begin{eqnarray}
\mathsf{\nabla }^{2}\mathcal{S} &=&-\mathcal{SV}  \label{e6} \\
\mathsf{\nabla }^{2}\mathcal{V} &=&-\mathcal{S}^{2}  \label{e7} \\
\alpha ^{2}\int_{0}^{\infty }d^{D}\mathsf{x}\,\mathcal{S}^{2} &=&1
\label{newnorm}
\end{eqnarray}
where we can assume that $\psi $ is real without loss of generality. The
parameters $\hat{\alpha}$ and $\beta $ have units of $\left( \text{length}%
\right) ^{2-D/2}$ and $\left( \text{length}\right) ^{2}\times $ energy
respectively. \ The system (\ref{e6},\ref{e7}) is invariant under the
rescaling transformation 
\begin{equation}
\left( \mathcal{S},\mathcal{V},\mathsf{x}\right) \longrightarrow \left(
\lambda ^{2}S,\lambda ^{2}V,\lambda ^{-1}x\right)  \label{rescale}
\end{equation}
independent of the dimension $D$, where $\lambda $ has units of inverse
length. Using this transformation we can rewrite the system in terms of
fully dimensionless functions $\left( S,V\right) $ of dimensionless
variables. For the spherically symmetric case the $D$-dimensional Laplacian
operator is $\nabla ^{2}\varphi =r^{1-D}\left( r^{D-1}\varphi ^{\prime
}\right) ^{\prime }$ and so (\ref{e6},\ref{e7},\ref{newnorm}) become 
\begin{eqnarray}
\left( r^{D-1}S^{\prime }\right) ^{\prime } &=&-r^{D-1}SV  \label{e8} \\
\left( r^{D-1}V^{\prime }\right) ^{\prime } &=&-r^{D-1}S^{2}  \label{e9} \\
\alpha ^{2}\lambda ^{4-D}\int_{0}^{\infty }d^{D}x\,S^{2} &=&1
\label{normint}
\end{eqnarray}
where the prime denotes differentiation with respect to $r$. We must require
that the integral in (\ref{normint}) be finite in order for the SN\ system
to be physically meaningful. We therefore seek solutions to the SN\ system
that are finite for all values of $r\geq 0$; continuity requirements imply
that these solutions are everywhere smooth. Note that $\alpha
_{0}^{2}\lambda ^{4-D}$ is a dimensionless quantity.

However, our equation (\ref{normint}) clearly has a problem for $D=4$, where
we lose our ability to rescale using $\lambda $. Because we still require
the wavefunction to be normalized to unity, we are forced to introduce a
constant $\frak{K}>0$ in the $D=4$ case so that $S(r)\rightarrow \sqrt{\frak{%
K}}S(r)$. Then our equations (\ref{e8}), (\ref{e9}), and (\ref{normint})
become 
\begin{eqnarray}
\left( r^{3}S^{\prime }\right) ^{\prime } &=&-r^{3}SV  \label{e8D4} \\
\left( r^{3}V^{\prime }\right) ^{\prime } &=&-\frak{K}r^{3}S^{2}
\label{e9D4} \\
\alpha ^{2}\frak{K}\int_{0}^{\infty }d^{4}x\,S^{2} &=&1  \label{normintD4}
\end{eqnarray}
We will not always include this constant $\frak{K}$ explicitly in the
following discussion. It will simply be understood to occur in the rescaling
when $D=4$. Smoothness implies that $S^{\prime }(0)$ and $V^{\prime }(0)$
are finite. With this information we can rewrite (\ref{e8},\ref{e9}) as 
\begin{eqnarray}
S(r) &=&S_{0}+\frac{1}{2-D}\int_{0}^{r}x\left( 1-\left( \frac{x}{r}\right)
^{D-2}\right) S(x)V(x)dx  \label{e8a} \\
V(r) &=&V_{0}+\frac{1}{2-D}\int_{0}^{r}x\left( 1-\left( \frac{x}{r}\right)
^{D-2}\right) S^{2}(x)dx  \label{e9a}
\end{eqnarray}
It is straightforward to show that Picard's theorem \cite{Picard} is
satisfied by this system of equations, and so given $S_{0}$ and $V_{0}$ a
unique solution to (\ref{e8},\ref{e9}) exists within a range $[0,R\left(
S_{0},V_{0}\right) \,)$. The $D<3$ versions of the integral equations (\ref
{e8a},\ref{e9a}) require some care. Integrating (\ref{e8},\ref{e9})
explicitly for these cases yields 
\begin{eqnarray}
S(r) &=&S_{0}+\int_{0}^{r}\left( x-r\right) S(x)V(x)dx  \label{e8b} \\
V(r) &=&V_{0}+\int_{0}^{r}\left( x-r\right) S^{2}(x)dx  \label{e9b}
\end{eqnarray}
for $D=1$ and 
\begin{eqnarray}
S(r) &=&S_{0}+\int_{0}^{r}x\ln \left( \frac{x}{r}\right) S(x)V(x)dx
\label{e8c} \\
V(r) &=&V_{0}+\int_{0}^{r}x\ln \left( \frac{x}{r}\right) S^{2}(x)dx
\label{e9c}
\end{eqnarray}
for $D=2$ . Both of these sets of equations may be obtained formally from (%
\ref{e8a},\ref{e9a}) by inserting these values of $D$, the latter case being
understood as the limit $D\rightarrow 2$.

Note that for all of these cases the function $V(r)$ is montonically
decreasing since 
\begin{equation}
V^{\prime }(r)=-\int_{0}^{r}\left( \frac{x}{r}\right) ^{D-1}S^{2}(x)dx
\label{e10a}
\end{equation}
a result valid for all $D\geq 1$. Hence if $V_{0}\leq 0$, then $%
\lim_{r\rightarrow \infty }V(r)=-\infty $ and so $S(r)$ will also diverge
for large $r$. Consequently only $V_{0}>0$ is of physical interest. For $%
V_{0}>0$ the function $V(r)>0$ \ initially. By rewriting $S(r)=r^{\frac{2-D}{%
2}}s(r)$ equation (\ref{e8}) can be rewritten in the form 
\begin{equation}
r^{2}s^{\prime \prime }+rs^{\prime }+\left[ r^{2}V(r)-\left( \frac{D-2}{2}%
\right) ^{2}\right] s=0  \label{e11a}
\end{equation}
which for $V(r)=V_{0}$ is Bessel's equation. Hence near the origin we expect 
$s(r)$ to have oscillatory behaviour. However when $V(r)$ becomes negative
eq. (\ref{e11a}) is similar to the modified Bessel equation and the
behaviour of $s(r)$ will be a linear combination of exponentially amplified
and damped functions. Normalization requirements imply that only the
exponentially damped solutions are allowed. In this case $S(r)$ decays like
an exponential times an inverse power of $r$ and so the integral in (\ref
{e9a}) will be finite, yielding a finite $V(r)$ at large $r$. \ This
behaviour was already noted for the $D=3$ case \cite{moroz}; we see here
that it is valid for arbitrary $D>0.$

\section{Numerical Study of the D-dimensional SN\ Equations}

The form (\ref{e8},\ref{e9}) of the SN\ equations is not well-suited for
numerical study. \ A more convenient form is 
\begin{eqnarray}
\left( rS\right) ^{\prime \prime } &=&-rSV+\left( 3-D\right) S^{\prime }
\label{e11} \\
\left( rV\right) ^{\prime \prime } &=&-rS^{2}+\left( 3-D\right) V^{\prime }
\label{e12}
\end{eqnarray}
We are interested in finding solutions to the system (\ref{e11},\ref{e12})
for that are smooth and finite for all $r$. These are the bound state
solutions referred to earlier. We will require our initial conditions, $%
S(0)=S_{0}$ and $V(0)=V_{0}$ to be greater than zero, a constraint which
avoids both the trivial solutions $S=V=0$\ and the non-normalizable
solutions $V=\pm S=-2(D-4)r^{-2}$, and is consistent with the rescaling
freedom (\ref{rescale}). \ It is straightforward to obtain the power-series
solutions to (\ref{e11},\ref{e12}) \ near the origin: 
\begin{eqnarray}
S(r) &=&S_{0}-\frac{S_{0}V_{0}}{2D}r^{2}+\frac{S_{0}\left(
S_{0}^{2}+V_{0}^{2}\right) }{8D\left( D+2\right) }r^{4}-\frac{%
S_{0}V_{0}\left( \left( 5D+4\right) S_{0}^{2}+DV_{0}^{2}\right) }{%
48D^{2}\left( D+2\right) \left( D+4\right) }r^{6}+\cdots  \label{s-ser} \\
V(r) &=&V_{0}-\frac{S_{0}^{2}}{2D}r^{2}+\frac{S_{0}^{2}V_{0}}{4D\left(
D+2\right) }r^{4}-\frac{S_{0}^{2}\left( \left( 2D+2\right)
V_{0}^{2}+DS_{0}^{2}\right) }{24D^{2}\left( D+2\right) \left( D+4\right) }%
r^{6}+\cdots  \label{v-ser}
\end{eqnarray}
where the functions $S$ and $V$ and their first derivatives are required to
be finite at the origin. \ Using the rescaling freedom to set $V_{0}=1$
(which is equivalent to setting $\lambda ^{2}=V_{0}$), we see that for any
dimension $D$ the solutions near the origin depend only on a single free
parameter.

Integrating the system (\ref{e11},\ref{e12}) using a Fehlberg fourth-fifth
order Runge-Kutta method in MAPLE we find that in each dimension $D$ an
infinite set of discrete bound states appears as expected. These can be
identified by the number of local extrema. An infinite amount of precision
is required in the choice of $S_{0}$ such that the solutions do not diverge
as $r$ increases. This value of $S_{0}$ marks the transition between
solutions in which $S(r)$ diverges to $+\infty $ and solutions where $S(r)$
diverges to $-\infty $. As we increase the accuracy of $S_{0}$ for a
specific bound state, the value of $r$ at which $S(r)$ blows up increases.
We will see that the values for $S_{0}$ and the distance between bound
states vary significantly with dimension.

The general method for finding solutions to the $D$-dimensional system is
the same as that in the $D=3$ case \cite{moroz}. \ Choose a value of $%
S_{0}=S_{0}^{\left( n+1\right) +}$ for which $S(r)$\ has $n$ zeros and
diverges to $\left( -1\right) ^{n}\infty $ at finite $r$. \ Then select a
slightly smaller value of $S_{0}=S_{0}^{\left( n+1\right) -}$ which has $n+1$
zeros and diverges to $\left( -1\right) ^{n+1}\infty $ for some finite $r$.
\ Between these two values of $S_{0}$ is a value $S_{0}^{\left( n+1\right)
-}<\hat{S}_{0}^{(n+1)}<S_{0}^{\left( n+1\right) +}$ for which $S(r)$\ has $n$
zeros and is smooth and finite for all values of $r$ \ and is
square-integrable. \ The potential $V(r)$\ will also be smooth and finite,
with finite energy eigenvalue $E_{n+1}$. By successively narrowing this
interval the bound-state value of $S_{0}$ can be achieved to any desired
accuracy. \ The bound state wavefunction with $n+1$ zeros may be obtained by
choosing another value $S_{0}^{(n+2)+}<S_{0}^{\left( n+1\right) -}$ and
repeating the procedure with $n\rightarrow n+1$. \ We find that the values $%
\hat{S}_{0}^{(n+1)}$ are rapidly decreasing functions of the dimension $D$,
and that the bound state solutions appear to accumulate at $D=6$. We are
limited by numerical accuracy in carrying out our investigations for
dimensions $D\geq 6$ due to this effect. A summary of this data is presented
in Table 1.

\begin{center}
\bigskip 
\begin{tabular}{ccccc}
$D$ & $\hat{S}_{0}^{1}$ & $\hat{S}_{0}^{2}$ & $\hat{S}_{0}^{3}$ & $\hat{S}%
_{0}^{4}$ \\ 
$1$ & $1.5583977884$ & $0.379904201$ & $0.2128374651$ & $0.1475990005$ \\ 
$2$ & $1.2134344293$ & $0.6482524055$ & $0.4937184140$ & $0.41447908$ \\ 
$3$ & $1.0886370794$ & $0.8264742841$ & $0.7442133785$ & $0.70014479$ \\ 
$4$ & $1.0327684253$ & $0.9309542414$ & $0.9053504436$ & $0.89422924$ \\ 
$5$ & $1.0081005592$ & $0.9825730584$ & $0.9795819080$ & $>0.978$ \\ 
$6$ & $1.0000000000$ & $1.0000000000$ & $1.0000000000$ & $1$%
\end{tabular}

{\small Table 1. Initial conditions }$\hat{S}_{0}^{(n+1)}${\small \ for the
first }$\left( n=0\right) ${\small \ to fourth }$\left( n=3\right) ${\small %
\ bound state wavefunctions, for one to six spatial dimensions. The
accumulation of the higher bound states for large dimensions results in the
decreasing accuracy of the lower right hand entries of the table.}
\end{center}

Figures 1,2 and 3 illustrate the solutions for $S$ for $D=1$, for the first,
second and third bound state transitions. Figures 4 to 8 illustrate the
first bound state transitions for dimensions $D=2$ to $D=6$. Figure 9
illustrates the accumulation effect at $D=6$. 
\begin{figure}[p]
\begin{center}
\epsfig{file=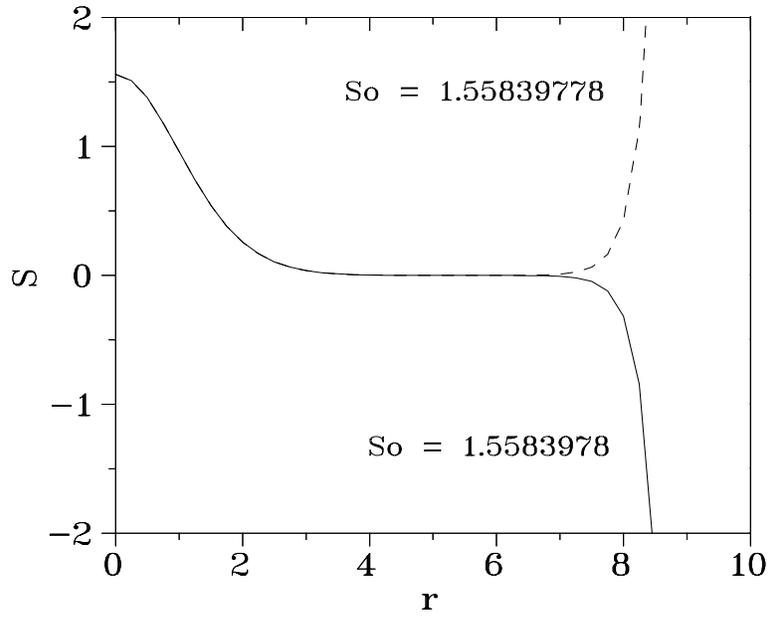, height=10cm, width = 8cm, angle = 90}
\end{center}
\caption{First bound state transition for the 1D SN equations}
\end{figure}
\begin{figure}[p]
\begin{center}
\epsfig{file=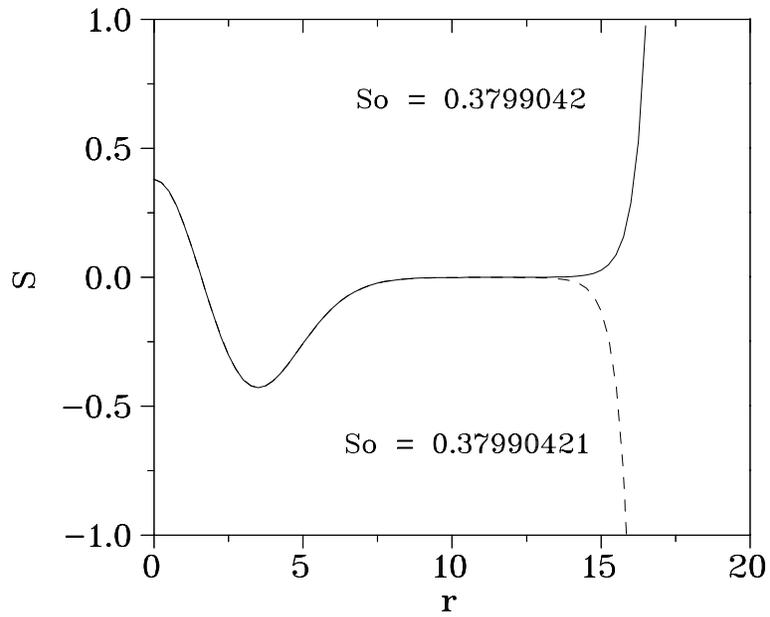, height=10cm, width = 8cm, angle = 90}
\end{center}
\caption{Second bound state transition for the 1D SN equations}
\end{figure}
\begin{figure}[p]
\begin{center}
\epsfig{file=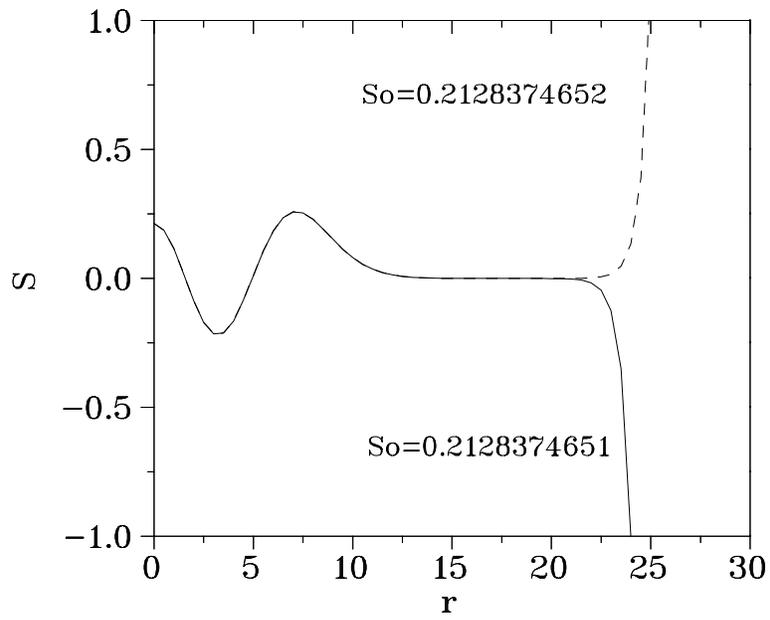, height=10cm, width = 8cm, angle = 90}
\end{center}
\caption{Third bound state transition for the 1D SN equations}
\end{figure}
\begin{figure}[p]
\begin{center}
\epsfig{file=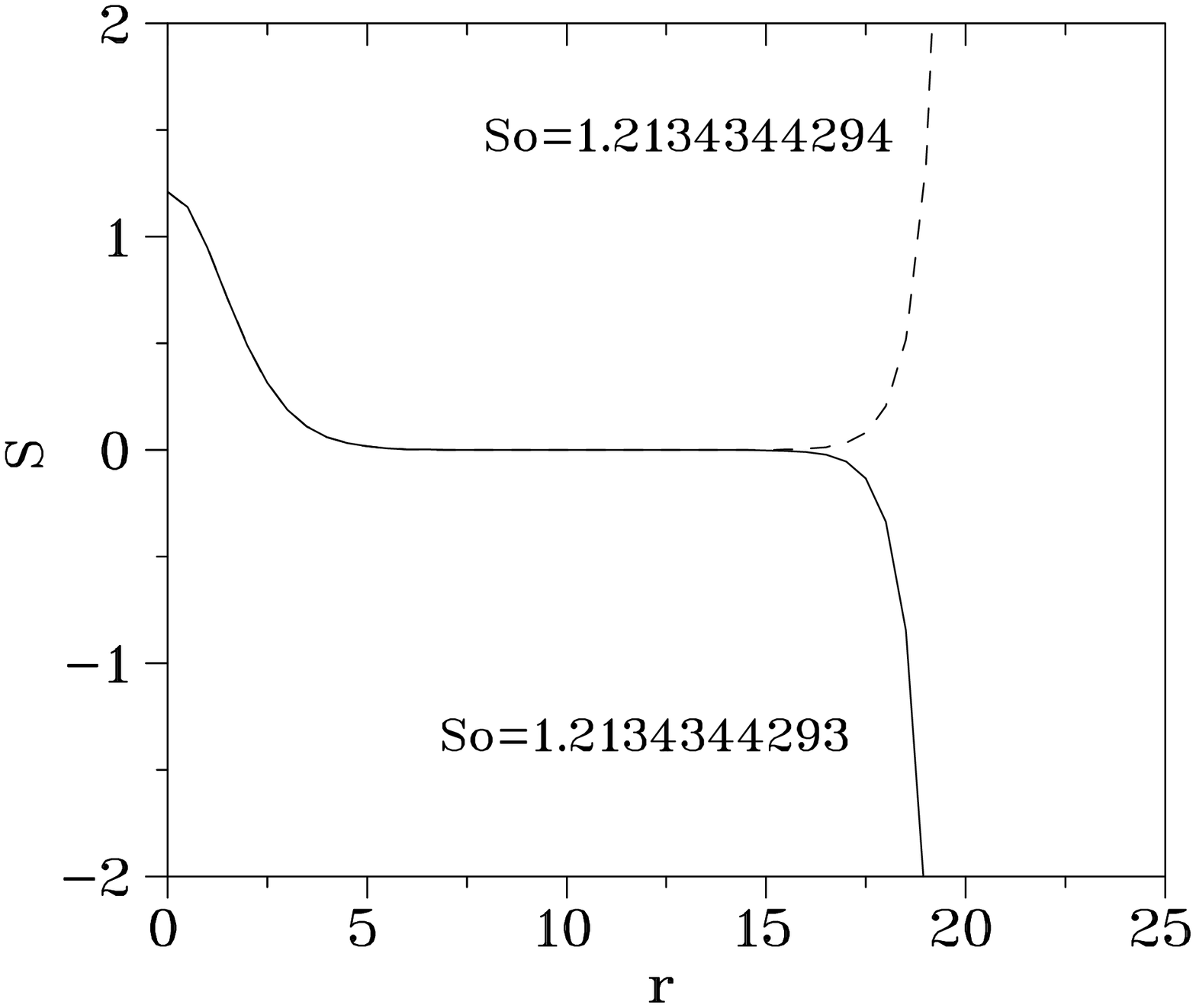, height=10cm, width = 8cm, angle = 90}
\end{center}
\caption{First bound state transition for the 2D SN equations}
\end{figure}
\begin{figure}[p]
\begin{center}
\epsfig{file=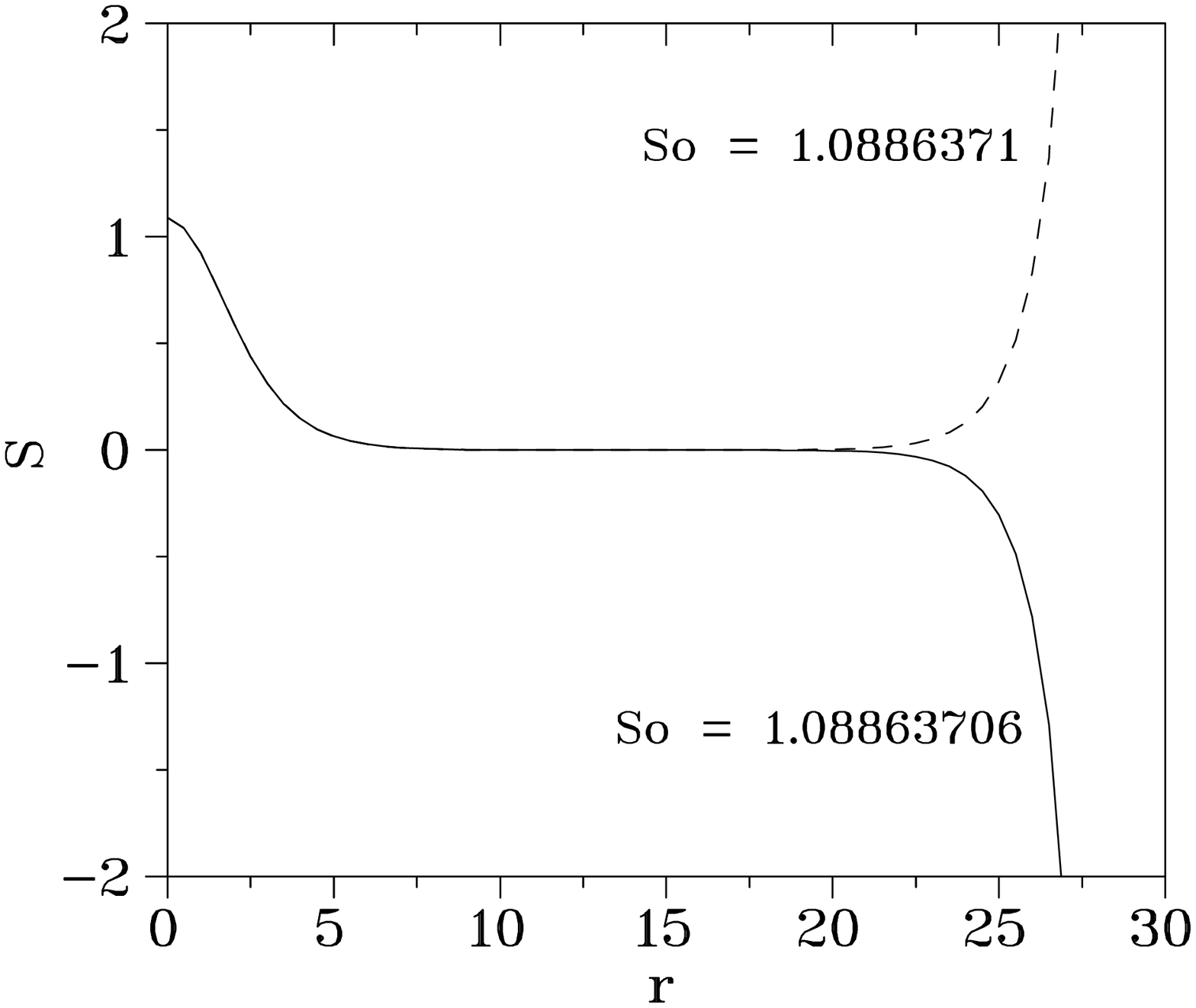, height=10cm, width = 8cm, angle = 90}
\end{center}
\caption{First bound state transition for the 3D SN equations}
\end{figure}
\begin{figure}[p]
\begin{center}
\epsfig{file=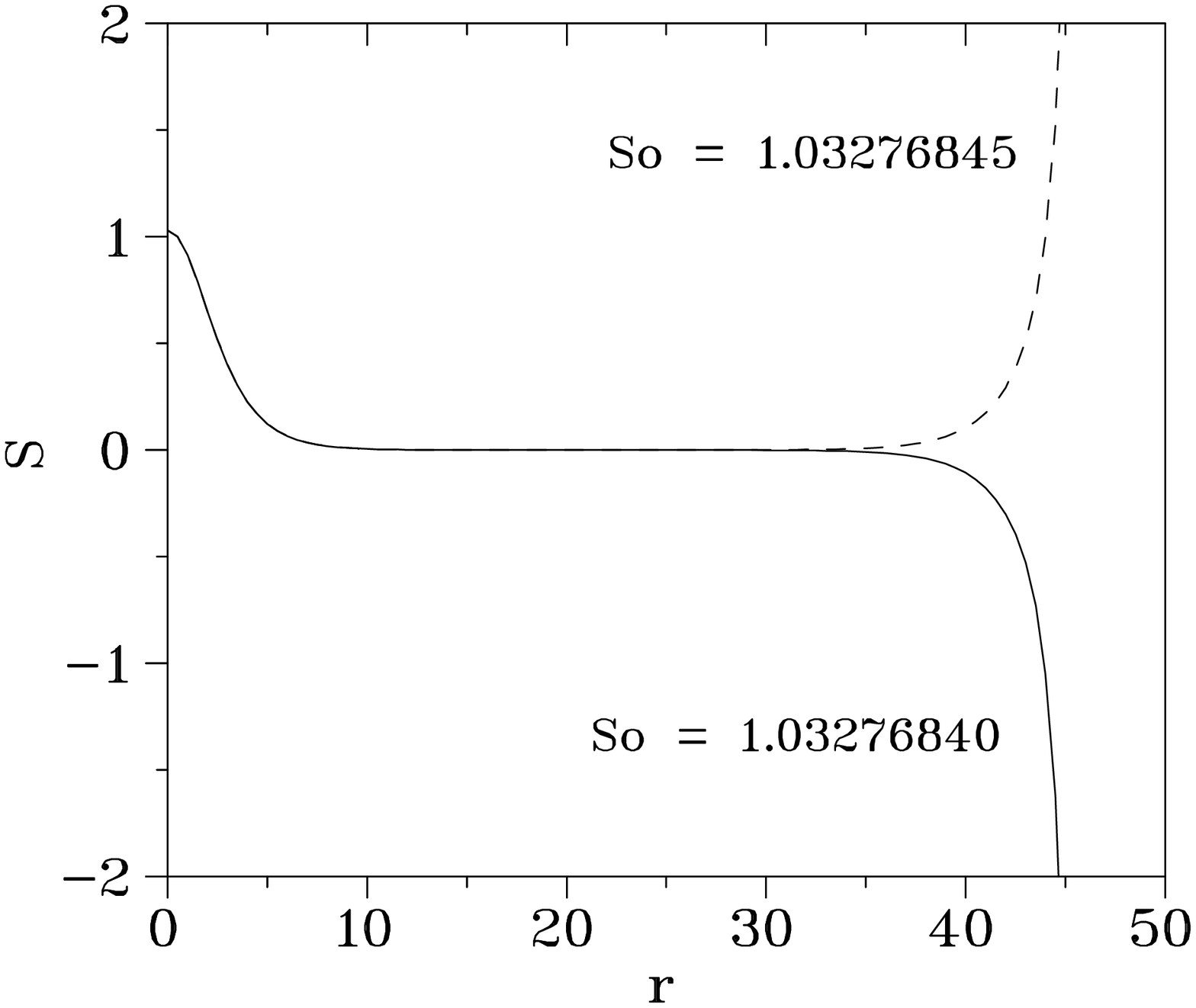, height=10cm, width = 8cm, angle = 90}
\end{center}
\caption{First bound state transition for the 4D SN equations}
\end{figure}
\begin{figure}[p]
\begin{center}
\epsfig{file=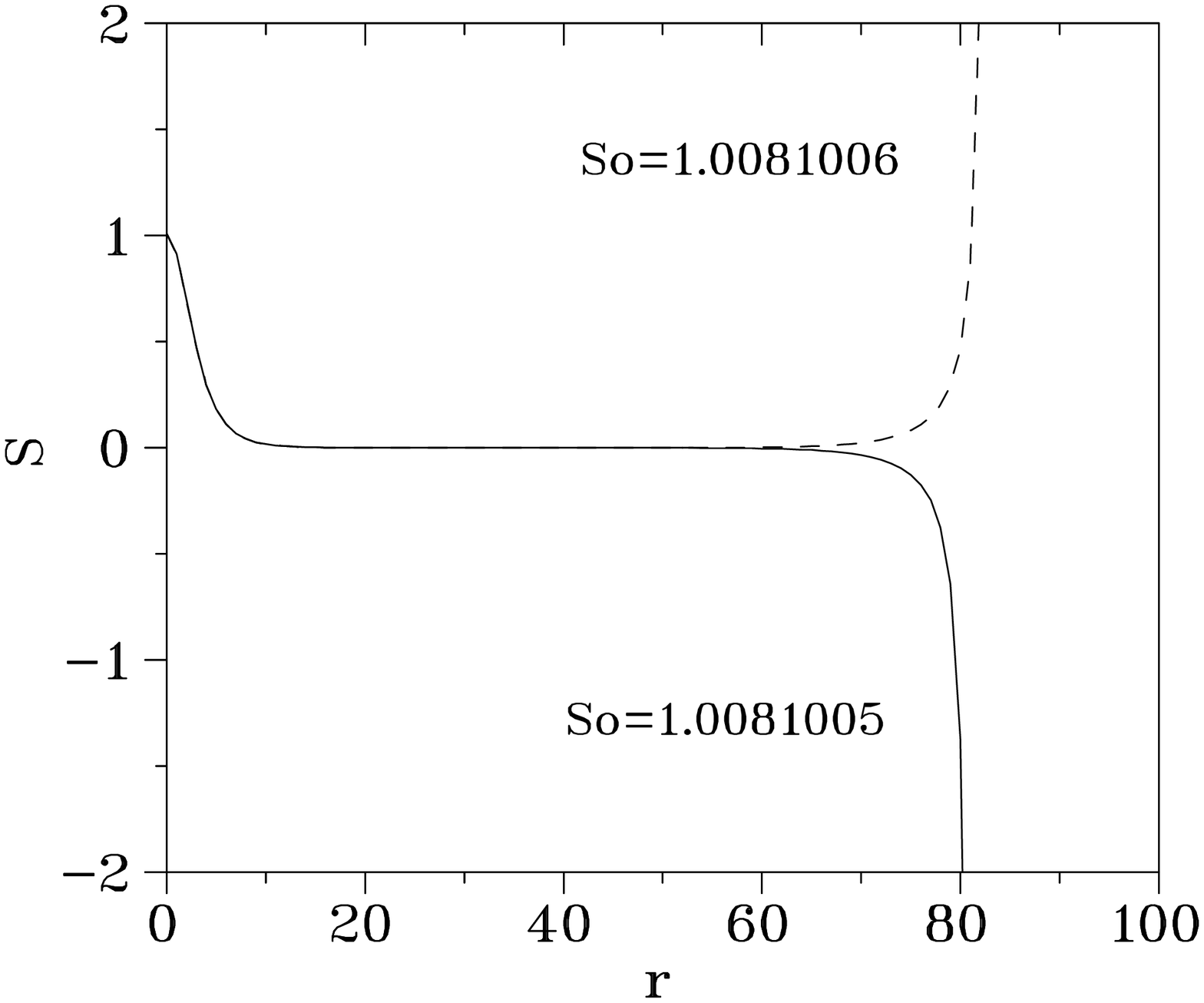, height=10cm, width = 8cm, angle = 90}
\end{center}
\caption{First bound state transition for the 5D SN equations}
\end{figure}
\begin{figure}[p]
\begin{center}
\epsfig{file=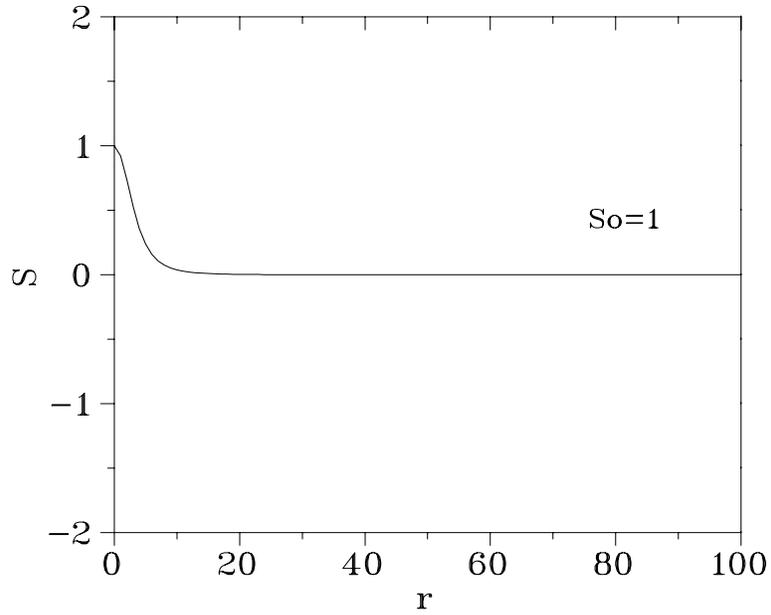, height=10cm, width = 8cm, angle = 90}
\end{center}
\caption{Integration of the 6D SN equations. Bound state characterictics at
this and higher dimensions are not apparent.}
\end{figure}
\begin{figure}[tbp]
\begin{center}
\epsfig{file=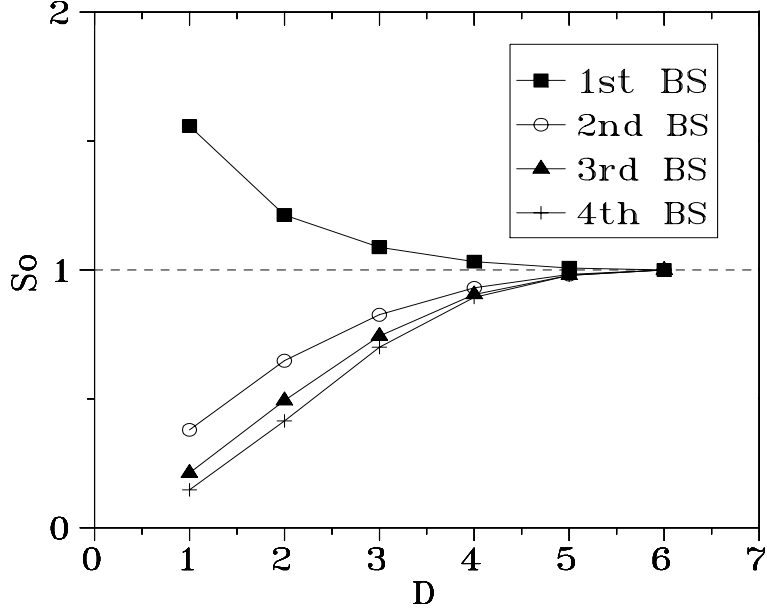, height=10cm, width = 8cm, angle = 90}
\end{center}
\caption{S(0) versus dimension. Bound states appear to accumulate at D = 6.}
\end{figure}
\bigskip

We turn now to the problem of \ determining the energy eigenvalues for the
bound states. From (\ref{e9}) and (\ref{e9a}) we find that the large-$%
\mathsf{r}$ expansion of $\mathcal{V}(\mathsf{r})$ is 
\begin{equation}
\mathcal{V}(r)=\mathcal{A}_{D}+\frac{\mathcal{B}_{D}}{D-2}\frac{1}{\mathsf{r}%
^{D-2}}+\cdots  \label{e13}
\end{equation}
where 
\begin{eqnarray}
\mathcal{A}_{D} &=&\mathcal{V}_{0}+\frac{1}{2-D}\int_{0}^{\infty }\mathsf{x}%
\mathcal{S}^{2}(x)d\mathsf{x}  \label{e14a} \\
\mathcal{B}_{D} &=&\int_{0}^{\infty }\mathsf{x}^{D-1}\mathcal{S}^{2}(\mathsf{%
x})d\mathsf{x}  \label{e14b}
\end{eqnarray}
These expansions are formally valid for all $D>0$; the explicit expansions
for $D<3$ are 
\begin{equation}
\mathcal{V}(\mathsf{r})=\left\{ 
\begin{array}{l}
\mathcal{A}_{1}-\mathcal{B}_{1}\mathsf{r}+\cdots \text{ \ \ \ \ \ \ \ \ \ \
\ \ \ \ \ \ \ }D=1 \\ 
\mathcal{A}_{2}-\mathcal{B}_{2}\ln \left( \frac{\mathsf{r}}{\mathsf{r}_{0}}%
\right) +\cdots \text{ \ \ \ \ \ \ }D=2
\end{array}
\right.  \label{e15}
\end{equation}
and 
\begin{eqnarray}
\mathcal{A}_{D} &=&\left\{ 
\begin{array}{l}
\mathcal{V}_{0}+\int_{0}^{\infty }\mathsf{x}\mathcal{S}^{2}(\mathsf{x})d%
\mathsf{x}\text{\ \ \ \ \ \ \ \ \ \ \ \ \ \ \ \ \ \ \ \ \ \ }D=1 \\ 
\mathcal{V}_{0}+\int_{0}^{\infty }\mathsf{x}\ln \left( \frac{\mathsf{x}}{%
\mathsf{r}_{0}}\right) \mathcal{S}^{2}(\mathsf{x})d\mathsf{x}\text{\ \ \ \ \
\ \ \ \ }D=2
\end{array}
\right.  \label{e16a} \\
\mathcal{B}_{D} &=&\left\{ 
\begin{array}{l}
\int_{0}^{\infty }\mathcal{S}^{2}(\mathsf{x})d\mathsf{x}\text{\ \ \ \ \ \ \
\ \ \ \ \ }D=1 \\ 
\int_{0}^{\infty }\mathsf{x}\mathcal{S}^{2}(\mathsf{x})d\mathsf{x}\text{\ \
\ \ \ \ \ \ \ \ }D=2
\end{array}
\right.  \label{e16b}
\end{eqnarray}
where $\mathsf{r}_{0}$ is an arbitrary length scale. \ The energy eigenvalue
is given by 
\begin{equation}
E_{D}=\beta \mathcal{V}\left( \mathsf{r}_{D}\right) =\beta \mathcal{A}_{D}
\label{E-eigen}
\end{equation}
The quantity $\mathsf{r}_{D}$ is the point at which the potential $U(\mathsf{%
r})$ vanishes. \ For $D\geq 3,$ $\mathsf{r}_{D}=\infty $. However for $D=1,2$
the potential diverges there and the situation is more delicate. \ For $D=1$
a Newtonian gravitational potential diverges linearly with $\mathsf{r}%
=\left| \mathsf{x}\right| $ . Here we extract the $r$-independent term from
( \ref{e15}) to obtain $E_{D=1}$; the result is still given by the
right-hand side of (\ref{e18}) with $D=1$. \ Effectively we have set $%
\mathsf{r}_{D}=0$ in (\ref{e15}) even though this equation is a large $%
\mathsf{r}$ expansion. \ For $D=2$ the Newtonian potential diverges at both
large and small $\mathsf{r}$. \ The normalization point $\mathsf{r}_{D}=%
\mathsf{r}_{0}$ is therefore arbitrary, and we set $E_{D=1}=\beta \lambda
_{D=2}^{2}A_{D=2}$ so that (\ref{e18}) remains valid in this case as well.
This is tantamount to setting $U(\mathsf{r}_{0})=\int_{\mathsf{r}%
_{0}}^{\infty }\mathsf{x}\ln \left( \frac{\mathsf{x}}{\mathsf{r}_{0}}\right)
S^{2}(\mathsf{x})d\mathsf{x}$ \ . A natural normalization point is to choose 
$\mathsf{r}_{0}$ to be the point at which the potential $V(\mathsf{r}_{0})=0$%
. \ \bigskip Note that under the rescaling transformation (\ref{rescale}) $%
\left( \mathcal{A},\mathcal{B}\right) \longrightarrow \left( \lambda
^{2}A,\lambda ^{4-D}B\right) $, where 
\begin{eqnarray}
A_{D} &=&1+\frac{1}{2-D}\int_{0}^{\infty }xS^{2}(x)dx  \label{e14c} \\
B_{D} &=&\int_{0}^{\infty }x^{D-1}S^{2}(x)dx  \label{e14d}
\end{eqnarray}
and so the combination $\mathcal{A}_{D}^{4-D}/\mathcal{B}%
_{D}^{2}=A_{D}^{4-D}/B_{D}^{2}$ is invariant. The parameter $\lambda $\ (or
alternatively $V_{0}$)\ is completely arbitrary, \ serving the sole function
of setting the length scales of the problem in units of $\alpha
_{0}^{2/\left( 4-D\right) }$. \ Solving (\ref{normint}) for $\lambda $
yields 
\begin{equation}
\lambda _{D}=\left( \frac{\Gamma \left( \frac{D}{2}\right) }{2\pi ^{\frac{D}{%
2}}\alpha ^{2}B_{D}}\right) ^{\frac{1}{4-D}}  \label{e17}
\end{equation}
With this normalization, we find that 
\begin{equation}
E_{D}=\beta \mathcal{A}_{D}=\beta \lambda _{D}^{2}A_{D}=\beta \left( \frac{%
\Gamma \left( \frac{D}{2}\right) }{2\pi ^{\frac{D}{2}}\alpha ^{2}}\right) ^{%
\frac{2}{4-D}}\frac{A_{D}}{\left( B_{D}\right) ^{\frac{2}{4-D}}}  \label{e18}
\end{equation}
and so the $D$-dimensional energy is given in units of $\beta /\alpha
^{4/(4-D)}$ . The quantities $A_{D}$ and $B_{D}$ \ are straightforwardly
solved numerically from (\ref{e14a},\ref{e14b}). The preceding expression
can be rewritten as 
\begin{equation}
E_{D}=\frac{1}{2}m_{pl}c^{2}\left( \frac{m}{m_{pl}}\right) ^{\frac{D+2}{4-D}%
}\left( \frac{4\Gamma \left( \frac{D}{2}\right) k^{2}}{\pi ^{\frac{D-2}{2}}}%
\right) ^{\frac{2}{4-D}}\frac{A_{D}}{\left( B_{D}\right) ^{\frac{2}{4-D}}}
\label{Eplanck}
\end{equation}
where $m_{pl}=\left( \hslash ^{D-2}/G_{D}c^{D-4}\right) ^{1/\left(
D-1\right) }$ is the Planck mass in $D>1$\ dimensions (in one spatial
dimension the quantity $c^{3}/\hslash G_{D}$\ is unitless, and can be
absorbed into the normalization constant $k$).

For $D\leq 3$ the energy eigenvalue is an increasing function of the
particle mass, whereas for $D\geq 5$ it is a decreasing function. \ The most
rapid increase is $E_{D}\sim m^{5}$ for $D=3$, with only linear and
quadratic dependence in $D=1$ and $2$ respectively. The time-scale for
collapse of the state vector is therefore most rapid in a world with three
spatial dimensions for any bodies whose mass is appreciably larger than the
Planck mass. \ In a world of more than four dimensions the collapse is
fastest for the lightest-mass particles, leading to behaviour which is at
complete odds with that expected in the macroscopic world.

\bigskip

If we interpret the lower-dimensional cases to be situations in three
spatial dimensions with planar or cylindrical symmetry, a somewhat different
picture emerges. Inserting the constants into (\ref{Eplanck}), the
dependence of the energy eigenvalues for the planar, cylindrical and
spherically symmetric cases are respectively 
\begin{eqnarray}
E_{p} &=&\frac{\left( 4\pi \right) ^{2/3}}{2}m_{pl}c^{2}\left( \frac{m}{%
m_{pl}}\right) \left( \frac{\ell _{pl}}{\ell }\right) ^{\frac{4}{3}}\frac{%
A_{1}}{\left( B_{1}\right) ^{\frac{2}{3}}}  \label{evplane} \\
E_{c} &=&2m_{pl}c^{2}\left( \frac{m}{m_{pl}}\right) ^{2}\left( \frac{\ell
_{pl}}{\ell }\right) \frac{A_{2}}{B_{2}}  \label{evcyl} \\
E_{s} &=&2m_{pl}c^{2}\left( \frac{m}{m_{pl}}\right) ^{5}\frac{A_{3}}{\left(
B_{3}\right) ^{2}}  \label{evsph}
\end{eqnarray}
where $\ell _{pl}$ is the Planck length and $\ell $ is the dimension of the
box (or cylinder) in which the particle is confined. \ From the perspective
of state-vector reduction, a particle prepared in a plane-wave or
cylindrical state will have a radically different time-scale of collapse
than the same particle prepared in a spherically symmetric state. A neutron
in a spherically symmetric state will have a collapse time of $\Delta t\sim
10^{53}s$ whereas a neutron confined to a square box $10$m in
cross-sectional size will have a collapse time of only $\Delta t\sim 10^{22}s
$. Both of these times are much longer than the age of the universe. However
by confining a neutron to a rectangular pipe whose width is on the order of
10 picometers, the collapse time becomes on the order of $10^{6}s$, or about
12 days. \ Such an experiment would push the limits of current technologly.

\bigskip

Note that for $D=4$ the probability is scale-invariant, and it is not
possible to solve (\ref{normint}) for $\lambda $. The energy scale in this
dimension is independent of the wavefunction normalization. \ If we work
through the problem using the $\frak{K}$ rescaled equations (\ref{e9D4}) and
(\ref{normintD4}), we can recover an expression for the energy eigenvalues
in this dimension using (\ref{E-eigen}). When rescaling equations (\ref{e13}%
) we can choose to absorb $\frak{K}$ in the rescaling of the expansion
coefficients: $\left( \mathcal{A},\mathcal{B}\right) \longrightarrow \left( 
\frak{K}\lambda ^{2}A,\frak{K}B\right) $ where $A$ and $B$ are equivalent to
(\ref{e14c}) and (\ref{e14d}). Then solving (\ref{normintD4}) for $\frak{K}$
we get

\begin{equation}
\frak{K} = \frac{\Gamma(2)} {2\pi ^{2}\alpha ^{2}B_{4}}  \label{SolK}
\end{equation}

We use this normalization to solve for the energy eigenvalue when $D=4$ 
\begin{equation}
E_{4}=\beta \mathcal{A}_{4}=\beta \frak{K}\lambda _{4}^{2}A_{4}\frak{=}%
\lambda _{4}^{2}\beta \left( \frac{\Gamma (2)}{2\pi ^{2}\alpha ^{2}}\right) 
\frac{A_{4}}{B_{4}}
\end{equation}
Clearly we are left with an unsolved parameter, $\lambda _{4}$, which is at
this point completely arbitrary. This is a consequence of the introduction
of the rescaling constant $\frak{K}$ in the $D=4$ case.

\bigskip

A graph of the invariant $A^{(4-D)}/B^{2}$ is instructive because it
reflects the behaviour of the energy eigenvalue solutions with bound state
that is independent of the value of the $D$-dimensional Newton constant.
Figure 10 shows that the energy eigenvalues are positive and increase with
bound state for $D<3$, consistent with what we expect for the form of the
classical Newtonian gravitational potential in these dimensionalities.
Figure 11 shows that the energy eigenvalues are negative for dimensions
greater than 3. In this graph, the values of $\lambda _{4}$ and $r_{0}$ are
set to unity for $D=4$.

\begin{figure}[p]
\begin{center}
\epsfig{file=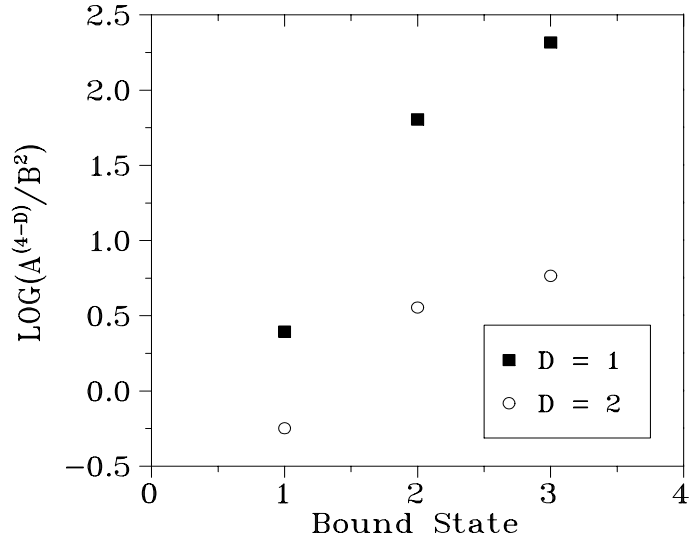, height=9cm, width = 7cm, angle = 90}
\end{center}
\caption{Behaviour of the invariant $A^{(4-D)}/B^{2}$ for the first three
bound states for D = 1 and 2}
\end{figure}

\begin{figure}[p]
\begin{center}
\epsfig{file=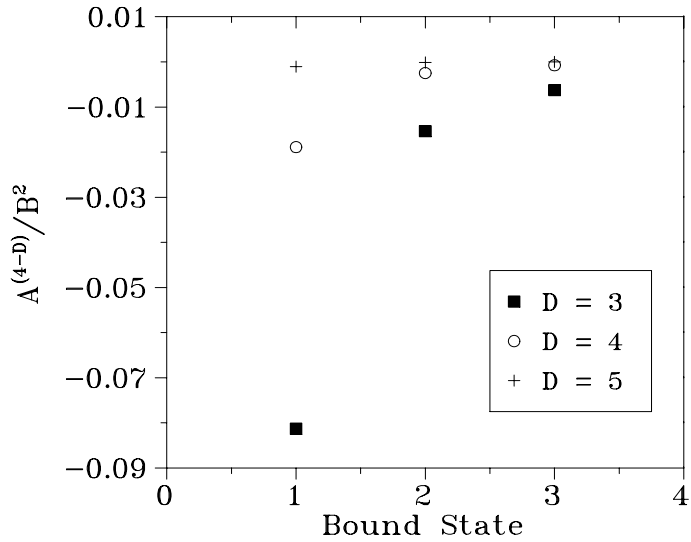, height=9cm, width = 7cm, angle = 90}
\end{center}
\caption{Behaviour of the invariant for the first three bound states for D =
3 to 5. In the four dimensional case, the invariant is $A_{4}/B_{4}$. The
value of $A^{(4-D)}/B^{2}$ for D = 6 is equal to zero within the accuracy of
our calculation}
\end{figure}

\section{Conclusions}

We have presented the preliminary results of our numerical and analytical
investigation of the Schroedinger-Newton equations in $D$ dimensions under
the assumption of spherical symmetry. Consistent with previous analyses in 3
dimensions, we have demonstrated numerically the existence of a discrete set
of \ ``bound-state'' solutions for the equations, which are associated with
different energy eigenvalues. The bound states appear to accumulate for
dimensions greater than or equal to six. The bound state energies for $D<3$
are positive and increase with increasing bound state number. For $D\geq 3$
the energies are negative and converge to zero with increasing eigenvalue
number. This is consistent with Moroz \cite{moroz}, who found that for $D=3$
the energy eigenvalues converge to zero as the order of the bound state
increases. For higher dimensions, the variation of the eigenvalues with
bound state becomes less pronounced, as illustrated in Figure 11. For $D>5$
we find that the eigenvalues accumulate, and we are unable to determine any
energy differences between bound states to 10 significant digits.

Moroz ventured further in the three dimensional study to assign numerical
estimations for timescales of collapse, using the lowest bound state
solutions to determine energy eigenvalues. However the analogous calculation
using general $D$ dimensions is not strictly correct due to possible
dimensional variation in Newton's constant $G_{D}$ which appears in (\ref{e5}%
). We will not attempt to predict the variation of $G_{D}$ with dimension
here, although we note that Kaluza-Klein theory suggests that $G_{D}$ is
proportional to the three dimensional $G_{3}$ divided by the square root of
the volume element of the extra space dimensions. We leave any such
numerical estimates for future work. \ 

However if we interpret the lower-dimensional cases to be systems in three
spatial dimensions with either planar or cylindrical symmetry, a different
picture emerges. The collapse time scale depends not only on the mass of the
particle but also on the size of the box in which it is confined. \ If the
gravitational influence of the box can be neglected and if the gravitational
field of the neutron in a plane wave state has planar symmetry, then we find
that the collapse timescales differ enormously between the two situations. \
A free neutron under its own gravitational self-influence will experience a
collapse of its wavefunction on a timescale many orders of magnitude larger
than the age of the universe, whereas one confined to a rectangular pipe
whose cross-sectional width is about 10 picometers in size will decay in
about two weeks. \ 

An obvious extension of our work involves introducing external field into
our version of the SN equations, in the form of a point source perturbation.
This can be added to the Schroedinger equation to give the perturbed system 
\begin{eqnarray}
-\frac{\hslash ^{2}}{2m}\nabla ^{2}\Psi +U\Psi -\frac{M}{r}\Psi &=&E\Psi
\label{e2PS} \\
\nabla ^{2}U &=&4\pi Gm^{2}\left| \Psi \right| ^{2}  \label{e3PS}
\end{eqnarray}
Here $M$ is the mass which generates the perturbing field. This system is
quite interesting in the respect that the form of the bound state solutions
is drastically altered from that of the ``free field'' equations considered
in this paper. A preliminary study of this system indicates that each bound
state solution seems to have three regions. For small $M$, the solutions
asymptotically approach those of the free field or non-perturbed case, as
expected. For moderate values of $M$, the equations are goverened by both
the free field and point source field, causing the values of $S_{0}$, for
which the solutions do not diverge, to alter considerably. Finally, for very
large $M$, the solutions are goverened almost entirely by the point source
perturbation, and become similar to the hydrogen atom solutions of the
Schroedinger equation. The effect of incresing dimension causes the
``transition'' values of $M$ (which separate the different regions of the
solutions) to decrease. A full study of the point source perturbation is
currently in progress. \bigskip

\textbf{Acknowledgments} This work was supported in part by the Natural
Sciences and Engineering Research Council of Canada. 

\end{document}